\documentclass[showpacs,preprintnumbers,amsmath,amssymb]{revtex4}

\usepackage{graphicx}

\def\bra#1{\left\langle #1 \right\vert}
\def\ket#1{\left\vert #1 \right\rangle}

\begin{document}
\title{Multi-qubit compensation sequences}

\author{Y Tomita, J T Merrill, and K R Brown}
\affiliation{School of Chemistry and Biochemistry and Computational Science and Engineering Division, Georgia Institute of Technology, Atlanta, Georgia 30332, USA}
\altaffiliation{ken.brown@chemistry.gatech.edu}

\begin{abstract}

The Hamiltonian control of $n$ qubits requires precision control of both the strength and timing of interactions. Compensation pulses relax the precision requirements by reducing unknown but systematic errors. Using composite pulse techniques designed for single qubits, we show that systematic errors for $n$ qubit systems can be corrected to arbitrary accuracy given either two non-commuting control Hamiltonians with identical systematic errors or one error-free control Hamiltonian.  We also examine composite pulses in the context of quantum computers controlled by two-qubit interactions.  For quantum computers based on the XY interaction, single-qubit composite pulse sequences naturally correct systematic errors. For quantum computers based on the Heisenberg  or exchange interaction,  the composite pulse sequences reduce the logical single-qubit gate errors but increase the errors for logical two-qubit gates.
 
\end{abstract}

\maketitle

\section{Introduction}

The control of quantum bits for quantum computation requires a high degree of accuracy. Aside from coherence and random noise, systematic errors limit our ability to control these quantum systems. These errors include slow fluctuations in control parameters relative to the experimental time and slight imperfections in fabrication. Overcoming these systematic errors will be crucial to achieve the potentially high-accuracy gates required for fault-tolerant quantum computation with local gates \cite{GottesmanJMO2000,SvoreQIC07,ClarkPRA2009}. The problem of unknown systematic errors has been studied extensively in NMR \cite{Freeman:book}. In NMR a large collection of spins are addressed by an RF field with an unknown spatial variation. To overcome this variation, broadband composites pulses were introduced \cite{Levitt1986,TyckoPRL1983}.

In principle, compensating pulses can be used to correct {\it unknown} systematic errors in single qubit gates to arbitrary order \cite{BrownPRA2004}. In a real experimental situation, other errors begin to accumulate and higher-order pulses may be of limited use \cite{XiaoPRA2006}.  The second order broadband pulse devised by Wimperis (BB1) \cite{WimperisJMR1994} is the standard of compensation and has been extended to two-qubit couplings by Jones \cite{JonesPRA2003}. In this paper, compensation pulses for multi-qubit systems and Hamiltonians are examined using BB1 as an example pulse.  BB1 and the higher-order pulse sequences of \cite{BrownPRA2004,AllwayJMR2007} are fully compensating; the pulses do not require a specific input state of the system and can be used to replace single pulses that are part of a larger sequence.

The paper is organised as follows: Section \ref{sec:control} describes how the control theory (and related geometry) of multiple qubits is suited for the type of compensation pulses used on single qubits. Section \ref{sec:notation} introduces our notation and a generalised BB1 sequence. Section \ref{sec:two-qubits} reexamines the two-qubit pulse sequence of Jones \cite{JonesPRA2003} in the case of multiple systematic errors. Different methods for creating BB1-style sequences are compared. Section \ref{sec:n-qubits} generalises to $n$ qubits and proves inductively that only two systematic errors need to be correlated to achieve arbitrary correction in all systematic errors.  Section \ref{sec:limuni} examines cases with sufficient control for universal quantum computation but not full control of the $n$ qubit space. Finally, we conclude in Section \ref{sec:conclusions}.

\section{Control theory and geometry of $n$ qubits}\label{sec:control}

The model we consider is $n$ qubits and $M$ dimensionless Hamiltonians denoted $H_m$. We define a pulse as applying $H_m$ with constant strengths $\Omega_m$ for a time $t$ where $\Omega_m$ is bound between $-\Omega_{\mathrm{max}}$ and $\Omega_{\mathrm{max}}$.  The resulting unitary evolution is $U(t)=\exp(-i\sum_m\Omega_m H_m t)$. The applied pulse may not create the desired evolution due to systematic errors in the control strength $\Omega_m^\prime=\Omega_m(1+\delta_m)$ and the timing $t^\prime=t(1+\delta_t)$. In this model, timing errors are correlated, while the individual strengths could have independent errors. The source of the errors will not be considered and we will examine unitaries of the form
$U(\{\theta\},\{\epsilon\})=\exp(-i\sum_m\theta_m(1+\epsilon_m)H_m)$.

The quantum system is universally controllable without unknown errors if $H_m$ generates the entire control algebra of su(2$^n$) by addition and the Lie bracket \cite{HuangJMP1983}. The very same technique can be used to determine if a composite pulse sequence exists \cite{LiPRA2006}. Additionally, the Lie bracket can be used to constructively build pulses, e.g.  the Solovay-Kitaev composite pulse sequences in \cite{BrownPRA2004}.

For $n$ qubits the corresponding Lie Algebra is su($2^n$). We choose as a convenient representation of the generators of the algebra, $\eta_j=\frac{1}{2}\bigotimes_{k=1}^{n}\sigma_{\lfloor (j~\mathrm{mod}~4^k) / 4^{k-1} \rfloor}$ where $\sigma_0=I$ is the identity on the qubit and $\sigma_{1}=X$, $\sigma_2=Y$, and $\sigma_3=Z$ are the single qubit Pauli operators.

There are $4^n-1$ operators since the generator of the global phase $\eta_0=\frac{1}{2}\bigotimes_{k=1}^n I$ is outside of the algebra of su($2^n$). For any two generators $\eta_i$ and $\eta_j$, we find that either they commute $[\eta_i,\eta_j]$=0 or $[\eta_i,\eta_j]=i\epsilon_{ijk} \eta_k$. If they do not commute, the two operators generate a representation of su(2).

The Lie algebra then imposes that given Pauli-operator generators with the same systematic control error, arbitrarily accurate composite pulses can be created, if and only if they do not commute. Furthermore, if they do not commute the resulting pulse sequence will have the same form as a single qubit pulse sequence \cite{LiPRA2006}. A geometrical interpretation is that controlling two elements that do not commute is homomorphic to rotations on a sphere while the space for commuting elements is a 2-torus \cite{ZhangPRA2003,KhanejaCP2001}. 

\section{Notation and BB1 revisited}\label{sec:notation}

The goal is to create accurate multi-qubit unitaries in the presence of systematic errors in $\theta_l$. For each case, we will start by defining the set of generators we control, $\{H_l\}$, and denote the unitary transformations as
\begin{eqnarray}
U_l(\theta_l)&=&\exp\left(-i \theta_l H_l\right) \nonumber \\
U_{l,m}(\theta_l,\theta_m)&=&\exp\left(-i (\theta_l H_l +\theta_m H_m)\right)\nonumber \\
U_{l,m,n}(\theta_l,\theta_m,\theta_n)&=&\exp\left(-i (\theta_l H_l +\theta_m H_m+\theta_n H_n)\right)\nonumber \\
U_{l,m,n,p}(\theta_l,\theta_m,\theta_n,\theta_p) &=&....
\end{eqnarray}

We will be particularly interested in sets of three generators that have commutation relations equivalent to su(2). In this case, rotations around the sphere can be used to guide the mathematics. The compensation pulses we present require that we can perform both positive and negative rotations. Physically this corresponds to inverting applied fields and changing the sign of multi-qubit interactions.

 A useful metric for evaluating the effects of control errors is the infidelity, $1-F(U,V)$, where $F$ is the fidelity,
 \begin{equation}
F(U,V)=\min_\psi \sqrt{\bra{\psi}U^\dagger V\ket{\psi}\bra{\psi}V^\dagger U\ket{\psi}},
\end{equation}
where $U$ is the ideal unitary and $V$ is the actual operation affected by the systematic error $\epsilon$.  We choose this measurement over the distance, $D(U,V)=\|U-V\|$, to avoid complications due to a global phase, {\it e.g.}, $U=X$ and $V=-X$. For $U=U_1(\theta_1)$ and $V=U_1(\theta_1(1+\epsilon_1))$, the distance scales as $O(\epsilon)$ and the infidelity scales as $O(\epsilon^2)$  \cite{DawsonQIC2006,GilchristPRA}.

Imagine we would like to perform $U_1(\theta)$ but our systematic control errors limit us to control of the form $U_{1,2}(\theta_1(1+\epsilon_1),\theta_2(1+\epsilon_2))$. Compensation sequences minimise the effect of these errors by applying successive error-prone pulses that cancel the leading error terms. In this notation, the BB1 sequence \cite{WimperisJMR1994} is
\begin{equation}
V_{W}(\theta,H_1,H_2)=U_1(\theta(1+\epsilon_1))T_{W}(\phi,H_1,H_2), 
\end{equation}
where $T_{W}(\phi,H_1,H_2)$ is the correction sequence with $\phi= \mathrm{acos}(-\theta/4 \pi)$
\begin{eqnarray}
T_{W}(\phi,H_1,H_2)=U_{1,2}(\pi\cos(\phi)(1+\epsilon_1),\pi\sin(\phi)(1+\epsilon_2))\nonumber\\
\times U_{1,2}(2\pi\cos(3\phi)(1+\epsilon_1),2\pi\sin(3\phi)(1+\epsilon_2))\nonumber\\
\times U_{1,2}(\pi\cos(\phi)(1+\epsilon_1),\pi\sin(\phi)(1+\epsilon_2)).
\end{eqnarray}

We refer to this sequence as BB1-W and when $\epsilon_1=\epsilon_2=\epsilon$ the sequence yields an infidelity that scales as $\epsilon^6$, $1-F(V_W(\theta,H_1,H_2),U_1(\theta))=O(\epsilon^6)$, or a distance that scales as $\epsilon^3$, details in \ref{App:A}.  An infidelity that scales as $\epsilon^{2n}$ corresponds to a distance that scales as $\epsilon^n$ \cite{BrownPRA2004}.  The fine control of the relative amplitude or phase $\phi$ allows for the correction; the compensation of higher order terms relies on increasingly finer control. 

The BB1 pulse sequence was derived in the context of single spins in NMR where $H_1=\frac{1}{2}X$ and $H_2=\frac{1}{2}Y$ \cite{WimperisJMR1994}. In many controlled quantum systems, the control occurs in a rotating frame and the difference between applying the generator $H_1$ or $\cos(\phi)H_1 +\sin(\phi)H_2$ is phase shifting the applied oscillating field relative to the rotating frame \cite{RakreungdetPRA2009}. As a result, for single qubit gates it is often reasonable to assume $\epsilon_1=\epsilon_2$. 

\section{Two qubits and multiple errors}\label{sec:two-qubits}

Jones applied BB1 to two qubit gates \cite{JonesPRA2003}. His construction assumes that the single qubit gates are without error. In the context of NMR, the natural two-qubit Hamiltonian is $H_1=\frac{1}{2}Z_1Z_2$. The error in the control of $H_1$ is unrelated to the error in $H_2=\frac{1}{2}X_1$, in this case no error. The direct application of BB1-W by simultaneous $H_1$ and $H_2$ pulses would fail to correct the errors.   However, the error free rotations about $X_1$ allows us to construct unitaries that are generated by $H_3=\frac{1}{2}Y_1Z_2$. Since the algebra is equivalent to rotations, we can use a $y$ ($H_2$) rotation to rotate the $x$ axis ($H_1$) to an axis in the $x-z$ plane ($c_1H_1+c_2H_2$) yielding 
\begin{equation}
U_2(\phi)U_1(\theta)U_2(-\phi)=U_{1,3}(\theta\cos(\phi),\theta\sin(-\phi)).
\end{equation}

This identity was used by Jones \cite{JonesPRA2003} to create an alternative pulse sequence, we will refer to as BB1-J. BB1-J transforms the requirement of relative amplitude-control (BB1-W) into  the accurate control of a rotation. We note that in NMR the sign of the ZZ Hamiltonian $H_1$ is determined by the molecule \cite{Levitt1986}. This shows that not all of the control Hamiltonians require invertible couplings in order to compensate. The correction sequence is then
\begin{eqnarray}
V_{J}(\theta,H_1,H_2)= U_1(\theta(1+\epsilon_1))U_2(\phi(1+\epsilon_2))U_1(\pi(1+\epsilon_1))U_2^\dagger(\phi(1+\epsilon_2))\nonumber\\
\times U_2(3\phi(1+\epsilon_2))U_1(2\pi(1+\epsilon_1))U_2^\dagger(3\phi(1+\epsilon_2))\nonumber\\
\times U_2(\phi(1+\epsilon_2))U_1(\pi(1+\epsilon_1))U_2^\dagger(\phi(1+\epsilon_2)).\label{eqn:VJ}
\end{eqnarray}
This sequence yields an infidelity that scales as $\epsilon_1^6$ when $\epsilon_2=0$ \cite{JonesPRA2003}. The scaling for when $\epsilon_2\neq 0$ is examined in \ref{App:A}.  

The utility of the any fully-compensating pulse sequence is that it can be used to replace single pulses in a sequence. If $X_1$ and $Y_1$ have the same systematic error, we can correct the $X_1$ rotation by BB1-W before correcting the $Z_1Z_2$ transformation by BB1-J. The sequence of BB1-WJ is
\begin{eqnarray}
V_{WJ}(\theta,H_1,H_2,H_4)=U_1(\theta(1+\epsilon_1))V_{W}(\phi,H_2,H_4)U_1(\pi(1+\epsilon_1))V_{W}^\dagger(\phi,H_2,H_4)\nonumber\\
\times V_{W}(3\phi,H_2,H_4)U_1(2\pi(1+\epsilon_1))V_{W}^\dagger(3\phi,H_2,H_4)\nonumber\\
\times V_{W}(\phi,H_2,H_4)U_1(\pi(1+\epsilon_1))V_{W}^\dagger(\phi,H_2,H_4),
\label{eqn:VWJ}
\end{eqnarray}
where
\begin{equation}
V_{W}^\dagger(\phi,H_1,H_2)=T_{W,H_1,H_2}^\dagger( \mathrm{acos}(-\phi/4 \pi))U_1^\dagger(\phi(1+\epsilon_1)).
\end{equation}
This sequence replaces error prone $U_2(\phi(1+\epsilon_2))$ pulse with the corrected rotation $V_W$ generated by the BB1-W sequence.
Here, $\{H_1,H_2,H_3=-i[H_1,H_2]\}$ is a representation of su(2) and $\{H_2,H_4=\frac{1}{2}Y,H_5=-i[H_2,H_4]\}$ is also a representation of su(2). 
The assumption is that the errors of $H_2$ and $H_4$ are equivalent, $\epsilon_2=\epsilon_4$.
The infidelity then scales as $(\alpha \epsilon_1^3 +\beta \epsilon_1\epsilon_2^3)^2$ where $\alpha$ and $\beta$ are constants that depend on $\theta$, $H_1$, $H_2$, and $H_4$.

\begin{figure}
\begin{center}
\includegraphics[scale=0.8]{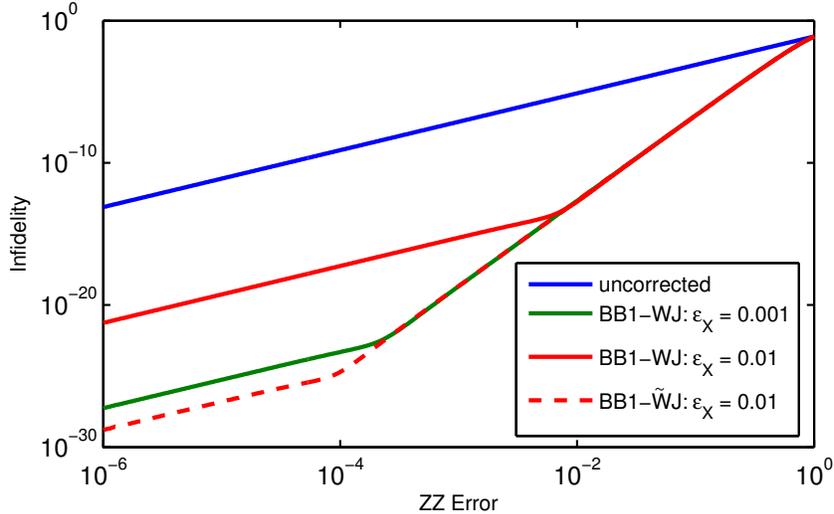}
\end{center}
\caption{Comparison of BB1-WJ and the higher order BB1-$\tilde{\mathrm{W}}$J pulse sequences applied to a $U_{ZZ}(\pi / 4)$ operation. For a fixed $X_1$ and $Y_1$ error $\epsilon_X$, the infidelity after a BB1-WJ correction scales as $(\alpha \epsilon_{ZZ}^3 +\beta \epsilon_{ZZ}\epsilon_{X}^3)^2$ (see text).  For the same $\epsilon_X$ the BB1-$\tilde{\mathrm{W}}$J sequence scales as $(\alpha \epsilon_{ZZ}^3 + \gamma_{4} \epsilon_{ZZ} \epsilon_{X}^{5})^2$, extending the regime where the infidelity scales as $\epsilon_{ZZ}^6$.}

\label{fig:compWJ}
\end{figure}

For fixed $\epsilon_2$, the infidelity at small $\epsilon_1$ scales as $\epsilon_1^2$ in $\epsilon_1$. This is the same order as the uncorrected pulse in $\epsilon_1$, although with a substantially smaller infidelity. In the case of $H_1=\frac{1}{2}Z_1Z_2$, $H_2=\frac{1}{2}X_1$, and $H_4=\frac{1}{2}Y_1$,  where $\epsilon_2=\epsilon_X=0.01$, the infidelity in this regime is a factor of $10^8$ smaller than the uncorrected pulse (see Figure \ref{fig:compWJ}). For $V_{WJ}$, the infidelity scales as $\epsilon_1^2$ when $\epsilon_1<\frac{\beta^2}{\alpha^2}\epsilon_2^{3/2}$. However, we can replace the $\mathrm{BB1-W}$ sequences $V_W$ in $V_{WJ}$ with higher order pulse sequences, for example the B$n$ sequences where B$2$=BB1 \cite{BrownPRA2004}. In this case, the infidelity will scale as $(\alpha \epsilon_1^3 +\gamma_n \epsilon_1\epsilon_2^{(n+1)})^2$, where $\gamma_n$ is a constant that depends on $\theta$ and B$n$. As a result, the value of $\epsilon_1$ where the scaling changes from  $\epsilon_1^6$ to $\epsilon_1^2$ becomes smaller and smaller. In Figure \ref{fig:compWJ}, we compare the scaling properties of the BB1-WJ and the higher order BB1-$\tilde{\mbox{W}}$J where we have replaced the $V_W$  BB1 sequence with the B4 sequence \cite{BrownPRA2004,XiaoPRA2006}. As expected, the error $\epsilon_1=\epsilon_{ZZ}$ where the scaling changes from $\epsilon^6$ to $\epsilon^2$ changes from $\approx 10^{-2}$ for BB1-WJ to $\approx 10^{-4}$ for BB1-$\tilde{\mbox{W}}$J.  In principle, given a target infidelity and systematic errors $\epsilon < 1$ \cite{BrownPRA2004}, we can construct a pulse sequence with an infidelity guaranteed below the target infidelity. We note that in practice other errors including random control errors and decoherence typically limit the fidelity. 

These sequences are each optimised for different correlations in the errors.  BB1-W performs well when errors in the control of $Z_{1}Z_{2}$ and $X_{1}$ are correlated while BB1-J is optimised for when one control has no error. BB1-WJ combines both strategies by first correcting the correlated errors and then correcting the independent error. 

\begin{figure}
\begin{center}
\includegraphics[scale=0.32]{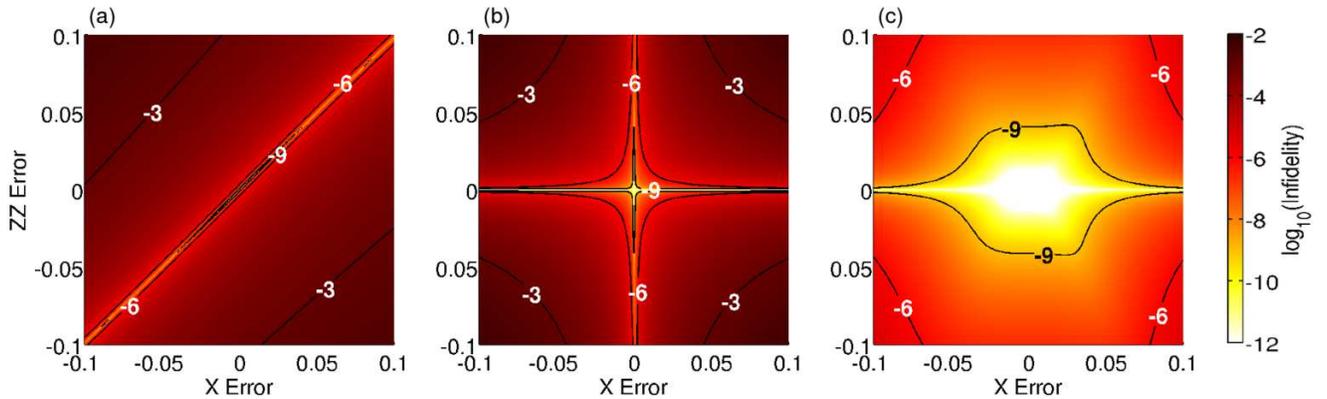}
\end{center}
\caption{Comparison of (a) BB1-W, (b) BB1-J, and (c) BB1-WJ pulse sequences applied on $U_{ZZ}(\pi / 4)$ operation on a pair of qubits. BB1-WJ assumes $X_1$ and $Y_1$ have equivalent systematic errors. 
}

\label{fig:compbb1}
\end{figure}

In Figure \ref{fig:compbb1}, we compare the ideal unitary $U=U_{ZZ}(\pi / 4)=\exp(-i \frac{\theta}{8} Z_1Z_2)$ to the approximate unitaries $V$ assuming errors equivalent errors in $X_1,Y_1$ and uncorrelated errors in $Z_{1}Z_{2}$. BB1-J ($V=V_J(\pi/4,Z_1Z_2/2,X_1/2)$) outperforms BB1-W ($V=V_W(\pi/4,Z_1Z_2/2,X_1/2)$) when either error is low. BB1-W is preferable when the systematic errors are identical. BB1-WJ ($V=V_{WJ}(\pi/4,Z_1Z_2/2,X_1/2,Y_1/2)$) results in low errors over the range of two errors. Initial compensation of the $X_1$ pulses results in better compensation of $Z_1Z_2$. 

\section{Extension to many qubits}\label{sec:n-qubits}

Given a control operator with a systematic error and a perfect rotation that transforms that operator to an orthogonal independent operator, we can perform compensation, {\it e.g.} BB1-J. Given two control operators with correlated errors that are generators of su(2), we can perform compensation, {\it e.g.} BB1-W. As a result, in principle one can perform arbitrarily accurate composite pulses on a controllable quantum system where all the controls have independent errors except two.  

As an example, imagine $n$ qubits in a row with single qubit operators and tunable Ising couplings. The Hamiltonians are $X_j$, $Y_j$ on each qubit and $Z_jZ_{j+1}$ between neighbours.  If for the qubit $n$, $X_n$ and $Y_n$ have uncorrelated error, there does not exist a compensation pulse \cite{LiPRA2006}. However, if the $X$ and $Y$ systematic errors are correlated on the the first qubit but otherwise independent, the following sequence can be used to generate an arbitrarily accurate $X$ rotation on the $n$th qubit.

For the initial qubit with correlated $X_1$ and $Y_1$ errors, BB1-W is used. To correct $Z_1Z_2$, BB1-J is used with BB1-W corrected $X_1$ pulses. This is the sequence BB1-WJ. $X_2$ on the second qubit is then corrected via BB1-J using BB1-WJ corrected $Z_1Z_2$ pulses. We denote this sequence as BB1-WJJ or BB1-WJ$^2$. Errors on the $n$th qubit can be compensated by repeated use of BB1-J along the chain, first correcting $X_j$, then $Z_jZ_{j+1}$ and then $X_{j+1}$ until $X_n$ is reached. The total sequence correcting the $n$th X rotation is denoted BB1-WJ$^{2(n-1)}$. 

\begin{figure}
\begin{center}
\includegraphics[scale=0.8]{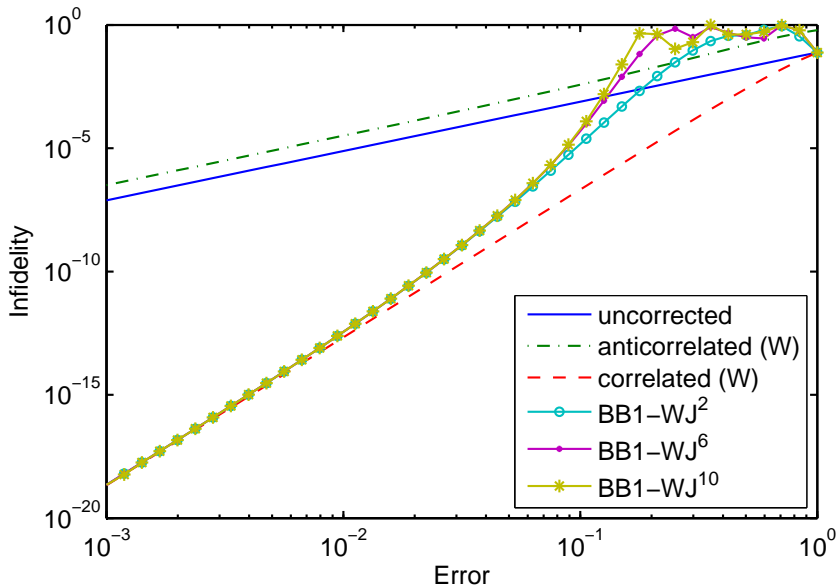}
\end{center}
\caption{Compensation of $U_{X_n}(\pi / 4)$ by application of BB1-WJ$^{2(n-1)}$. Compensation of $U_{X_n}$ by BB1-W pulses using $Y_n$ works only when the errors are correlated. Anticorrelated errors between $X_n$ and $Y_n$ increase the infidelity. BB1-WJ$^{2(n-1)}$ uses the correlated errors of $X_1$ and $Y_1$ and a chain of $Z_jZ_{j+1}$ interactions to compensate the $X_n$ rotation.  The results for $X_2$, $X_4$ and $X_6$ are shown.}
\label{fig:manyq}
\vspace{-1em}
\end{figure}

Figure \ref{fig:manyq} compares correcting a $\pi/4$ $X$ rotation as a function of chain length assuming equal magnitude errors for all operators but with a random sign except for $X_1$ and $Y_1$. The correlated and anti-correlated lines serve as references. If $X_n$ and $Y_n$ have correlated errors, then local BB1-W greatly reduces the infidelity. In the worst case scenario, the errors are anticorrelated and the compensation pulses add additional error to the initial overrotation. $X_n$ rotations can still be corrected using BB1-WJ$^{2(n-1)}$, if only $X_1$ and $Y_1$ are correlated.  The error increases with position (comparing BB1-WJ$^2$ to BB1-WJ$^{10}$) on the chain for large errors but approaches an equivalent fidelity for small errors. Asymptotically, the correction of $X_n$ rotations by sequential correction (BB1-WJ$^{2(n-1)}$) is equivalent to the BB1-W correction composed of correlated $X_n$ and $Y_n$ rotations. Replacing BB1 with the pulse sequences from \cite{BrownPRA2004} allows for the creation of arbitrarily accurate pulse sequences.

Although, this is not practical on a large scale, it can lead to a constant reduction in the number of gates that need to be calibrated at the beginning of an experiment for a large quantum system. Per region of computation, only a few highly reliable quantum gates can be used to reduce systematic errors in their neighbours. 

\section{Limited universality}\label{sec:limuni}
An interesting theoretical proposal with potential applications for quantum dots \cite{WeinsteinPRA2005,WeinsteinPRA2005-Levy}, superconducting qubits \cite{StorczPRB2005}, and trapped ions \cite{BrownPRA2003} is the use of only two-qubit interactions for quantum computation \cite{KempeQIC2001}. These two-qubit interactions are chosen to generate a sufficiently large algebra to create universal computation on a subspace of the total Hilbert space. We examine composite pulses for   $XY$ and Heisenberg interaction based quantum computers. The pulse sequences require that the sign of the two-qubit interactions can be inverted. Coupled quantum dots have been shown to exhibit reversible exchange couplings which can be controlled by an external magnetic field \cite{Burkard1999,Zumbuhl2004}, and may be promising candidates for this encoding.

\subsection{XY}

A Hamiltonian made of $XY$ interactions, $A_{\{i,j\}}=\frac{1}{2}(X_iX_j+Y_iY_j)$, has been shown to be universal over 3-qubits encoded into one-qubit \cite{KempeQIC2001,KempePRA2002}. The $XY$ interaction preserves the projection of angular momentum along $z$ but does not preserve total angular momentum. The qubit is encoded in a subspace of the qutrit defined by $m_z=1/2$ or $m_z=-1/2$.

\begin{figure}
\begin{center}
\includegraphics[scale=0.8]{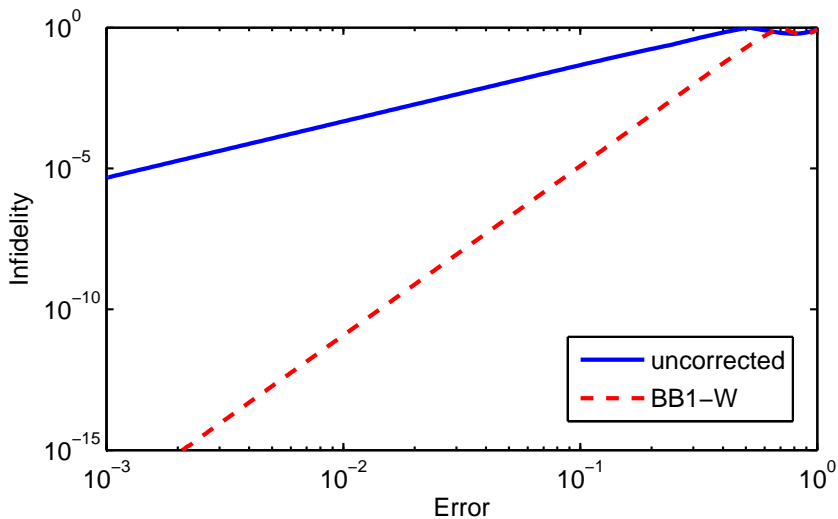}
\end{center}
\caption{Application of BB1-W pulse sequence correction on an effective $U_{\bar{Z}}(\pi / 4)$ operation ($T$ gate) on the $XY$ qubit following \cite{KempePRA2002}. $\bar{Z}$ rotations are implemented with the five pulse $\mathcal{P}_{3}$ sequence. Each pulse in the sequence is replaced with a BB1-W compensating pulse sequence.}
\label{fig:XY}
\vspace{-1em}
\end{figure}

Irrespective of the encoding, compensation is possible if the systematic errors are shared because
\begin{eqnarray}
A_{\{1,2\}}&=&\frac{1}{2}\left(X_1X_2+Y_1Y_2\right)\\
A_{\{2,3\}}&=&\frac{1}{2}\left(X_2X_3+Y_2Y_3\right)\\
A'=[A_{\{1,2\}},A_{\{2,3\}}]/(i)&=&\frac{1}{2}\left(X_1Z_2Y_3-Y_1Z_2X_3\right)
\end{eqnarray}
is a representation of su(2). For three qubits, we can block diagonalise the operators into four irreducible representations. For $m_z=\pm 3/2$, the irreducible representation is one-dimensional. For $m_z=\pm 1/2$ the irreducible representation is three-dimensional.

The operators $\bar{X}$, $\bar{Y}$ and $\bar{Z}$ act as the Pauli matrices on the encoded space and can be performed with pulses utilising the $XY$ interaction alone. For example, $\bar{Z}$ rotations on the encoded qubit are implemented by the five pulse $\mathcal{P}_3$ sequence which uses $XY$ interactions between each of the physical qubit pairs \cite{KempePRA2002,LidarPRL2001}.
\begin{eqnarray}
\mathcal{P}_3(\theta,\epsilon) &= U_{\{1,2\}}(-\frac{\pi}{4}(1+\epsilon)) U_{\{2,3\}}(-\frac{\pi}{2}(1+\epsilon)) U_{\{1,3\}}(-\frac{\theta}{2}(1+\epsilon))\nonumber\\
&\times U_{\{2,3\}}(\frac{\pi}{2}(1+\epsilon)) U_{\{1,2\}}(\frac{\pi}{4}(1+\epsilon))
\end{eqnarray}
When $\epsilon=0$, $\mathcal{P}_3(\theta,\epsilon)$ is equivalent to a rotation about the $z$ axis in the code space, $U_{\bar{Z}}(\theta)$, up to a global phase.  The $\mathcal{P}_3$ sequence explicitly requires invertible couplings between physical qubits, which may limit the types of systems that an $XY$ computer can be built from. Assuming the errors are proportional for each $H_{\{i,j\}}$, we can correct the timing error using BB1-W for each pulse; each $U_{\{k,j\}}(\theta(1+\epsilon))$ is replaced by $V_W(\theta,A_{\{k,j\}},A_{\{j,l\neq k\}})$. The results of using the correction are shown in Figure \ref{fig:XY}.

The remarkable part of the $XY$ interaction is that the su(2) algebra of neighbouring $XY$ operators is independent of our choice of encoded qubit. The exact same methods can be used to compensate the two-qubit gate sequences. Furthermore, we can apply our results for the $ZZ$ chain from Section \ref{sec:n-qubits} to show that only two neighbouring $XY$ interactions need to have identical systematic errors.

\subsection{Heisenberg}

\begin{figure}
\begin{center}
\includegraphics[scale=0.8]{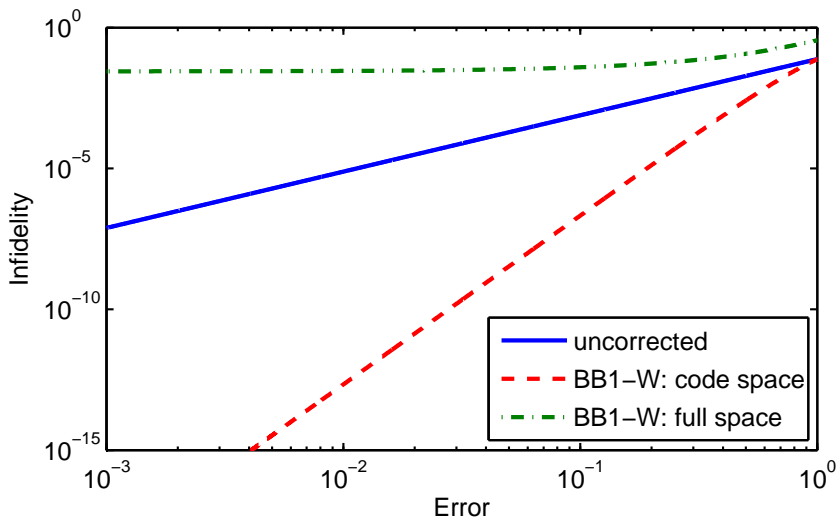}
\end{center}
\caption{Application of BB1-W pulse sequence correction for an $U_{\bar{Z}}(\pi/4)$ operation on a Heisenberg qubit encoded on three spins.  The qubit is encoded on a su(2) subspace on which compensation pulses are performed.  BB1 sequences greatly improve fidelity for states in the code space but fail for states with support outside the code space.}
\label{fig:heisenberg}
\end{figure}

The Heisenberg Hamiltonian $G_{\{i,j\}}=(X_iX_j+Y_iY_j+Z_iZ_j)$ has also been shown to be universal \cite{BaconPRL2000,KempePRA2001,DiVincenzoNature2000}. Furthermore, for certain arrangements of spins and exchanges it serves to protect errors by both energetics and symmetries \cite{BaconPRL2001}.  It is more convenient to write this as $G_{\{i,j\}}=(-I+2E(i,j))$ where $E(i,j)$ exchanges the states of qubits $i$ and $j$. As a result $[G_{\{i,j\}},G_{\{j,k\}}]=4(P(i,j,k)-P(i,k,j))$ where $P(i,j,k)$ is the cyclic permutation of $i$, $j$, and $k$. This does not result in a representation of su(2).

The exchange Hamiltonian $E(i,j)$ preserves both total angular momentum and the projection along $z$. For a single logical qubit made of three spins \cite{DiVincenzoNature2000}, any two exchange terms represent the algebra $\mathrm{u}(1) \bigoplus \mathrm{su}(2)$. The qubit is encoded into the su(2) block corresponding to the total angular momentum, $S=1/2$, and projection, $m_z=1/2$. The exchange Hamiltonian generates rotations equivalent to su(2) on the code space with $E(1,2)$ and $\frac{1}{\sqrt{3}} (E(1,2)+ 2E(2,3))$ corresponding to  $\bar{Z}$ and $\bar{X}$ respectively. Up to a global phase, $G_{\{i,j\}}$ and $2E(i,j)$ generate the same unitary evolution. Assuming that $G_{\{1,2\}}$ and $G_{\{2,3\}}$ have equivalent systematic errors and the interaction strengths can change sign, compensation is then possible on the code space using BB1-W.  

  It is not clear that the two-qubit gates can be corrected since the state has support on the u(1) and su(2) blocks. In Figure \ref{fig:heisenberg} we calculate the three-qubit fidelity after compensation. If we limit ourselves to states in the code space, the compensation works as expected. Allowing states outside of the code space, the compensating pulses are worse than the uncompensated pulse for low errors.  This result is expected since the operators do not form an su(2) algebra which the BB1 sequence depends on.

\section{Conclusions}\label{sec:conclusions}

We have shown that arbitrarily accurate compensation is possible with a fully controllable system if either two non-commuting Hamiltonians that generate su(2) have equivalent systematic errors or if a single Hamiltonian is error free. In the case of two non-commuting Hamiltonians with equivalent systematic errors, pulses of both positive and negative amplitudes are required.  The underlying pulse sequences are equivalent to sequences for qubits. Furthermore, we reemphasise the importance of the algebra and show that the same compensation pulses work for universal $XY$ quantum computation but not for universal Heisenberg quantum computation with a three-qubit encoding. 

Compensation pulses are well-suited for single qubits controlled by interaction with electromagnetic waves in the rotating frame. In this case, the difference between $X$ and $Y$ Hamiltonians is simply a change in the phase of the electromagnetic wave. Uncertainty in the amplitude of the applied field naturally leads to an unknown but equivalent error in the two Hamiltonians.
 
For systems based on two-qubit interactions, the requirement of positive and negative couplings can present a challenge. Furthermore for solid-state systems, the gate couplings will most likely have independent systematic errors. Our work presents a possible solution. For the XY case a single, invertible error-free XY interaction can be used to build accurate gates. Using the identity that $\exp(-i\pi A_{\{1,2\}})A_{\{2,3\}}\exp(i\pi A_{\{1,2\}})=-A_{\{2,3\}}$, we can create effective negative couplings for neighbouring XY interactions. We can then generate arbitrarily accurate unitary gates locally using BB1-J type sequences. These can interact with their neighbours, etc. This is impractical but it does suggest that a system with a few low-error invertible couplings could efficiently compensate neighbouring high-error couplings.  

The su(2) algebra underlying  these compensating pulse provides additional incentive to continue development of single qubit compensation pulses. Shaped pulse sequences or continuous time control can lead to further improvements  \cite{KhanejaJMR2005}. The question remains how to develop composite pulses that do not rely on a su(2) or so(3) subalgebra.  The Lie algebraic technique of \cite{LiPRA2006} rules out composite pulses with Heisenberg coupling. However, we know that over an encoded space at least space single qubit compensation pulses are possible.  The development of compensation pulses that do not use the geometry of the sphere and the development of techniques for identifying compensation compatible subspaces are both interesting challenges. 

\begin{acknowledgments}
This work was supported by Georgia Tech and by IARPA through the Army Research Office Grant No. W911NF-08-1-0515. YT acknowledges the support of an Emerson Fellowship . JTM acknowledges the support of a Georgia Tech Presidential Fellowship and an Emerson-Williams Fellowship.
\end{acknowledgments}
\appendix
\section{Analytical Evaluation of Errors in BB1 and BB1-J}\label{App:A}

The original BB1 sequence \cite{WimperisJMR1994} is BB1-W where both errors are equivalent
\begin{equation}
V(\theta,H_1,H_2)=U_1(\theta(1+\epsilon))T(\phi,H_1,H_2), 
\end{equation}
where $T(\phi,H_1,H_2)$ is the correction sequence with $\phi= \mathrm{acos}(-\theta/4 \pi)$
\begin{eqnarray}
T(\phi,H_1,H_2)=U_{1,2}(\pi\cos(\phi)(1+\epsilon),\pi\sin(\phi)(1+\epsilon))\nonumber\\
\times U_{1,2}(2\pi\cos(3\phi)(1+\epsilon),2\pi\sin(3\phi)(1+\epsilon))\nonumber\\
\times U_{1,2}(\pi\cos(\phi)(1+\epsilon),\pi\sin(\phi)(1+\epsilon)).
\end{eqnarray}
The $\pi$ rotations {\em toggle} $3\phi$ to -$\phi$ (see \cite{WimperisJMR1994}). After removing identities, we can rewrite $T(\phi,H_1,H_2)$ as
\begin{eqnarray}
T(\phi,H_1,H_2)=U_{1,2}(\pi\epsilon\cos(\phi),\pi\epsilon\sin(\phi))\nonumber\\
\times U_{1,2}(2\pi\epsilon\cos(\phi),-2\pi\epsilon\sin(\phi))\nonumber\\
\times U_{1,2}(\pi\epsilon\cos(\phi),\pi\epsilon\sin(\phi)).
\end{eqnarray}  
We use the Magnus expansion \cite{MagnusCPAM1954} to combine the three unitary operators, 
\begin{eqnarray}
T(\phi,H_1,H_2)&=&\exp\left(i \epsilon\theta H_1 + i \epsilon^3 M_3(\phi,H_1,H_2) +O(\epsilon^5)\right)\nonumber\\
&=&U_1(-\epsilon\theta)\left[1+i\epsilon^3M_3(\phi,H_1,H_2)+O(\epsilon^4)\right]
\end{eqnarray}  
where 
\begin{eqnarray}
M_3(\phi,H_1,H_2)=\frac{2\pi^3}{3}\cos(\phi)\sin^2(\phi)H_1+2\pi^3\cos^2(\phi)\sin(\phi)H_2.
\end{eqnarray}
The second order term vanishes due to the symmetry of the pulse sequence \cite{WimperisJMR1994,BrownPRA2004}.
As a result,
\begin{eqnarray}
V(\theta,H_1,H_2)&=&U_1(\theta(1+\epsilon))U_1(-\epsilon\theta)\left[1+i\epsilon^3M_3(\phi,H_1,H_2)+O(\epsilon^4)\right] \nonumber\\
&=&U_1(\theta)\left[1+i\epsilon^3M_3(\phi,H_1,H_2)+O(\epsilon^4)\right]. 
\end{eqnarray}
This shows that $V(\theta,H_1,H_2)$ is an approximation of $U_1(\theta)$ that scales as $\epsilon^3$ in distance \cite{BrownPRA2004} and $\epsilon^6$ in infidelity \cite{JonesPRA2003}. Both the distance and infidelity depend on the specific $H_1$ and $H_2$  
We perform a similar analysis for BB1-J. Starting from Equation \ref{eqn:VJ}, we find
\begin{equation}
 V_J(\theta,H_1,H_2)=U_1(\theta(1+\epsilon_1))T_J(\Phi,H_1,H_2), 
\end{equation}
where $T_J(\Phi,H_1,H_2)$ is the correction sequence with $\Phi= \mathrm{acos}(\theta/4 \pi)(1+\epsilon_2)=\phi(1+\epsilon_2)$
\begin{eqnarray}
 T_J(\phi,H_1,H_2)=U_{1,3}(\pi\cos(\Phi)(1+\epsilon_1),-\pi\sin(\Phi)(1+\epsilon_1))\nonumber\\
\times U_{1,3}(2\pi\cos(3\Phi)(1+\epsilon_1),-2\pi\sin(3\Phi)(1+\epsilon_1))\nonumber\\
\times U_{1,3}(\pi\cos(\Phi)(1+\epsilon_1),-\pi\sin(\Phi)(1+\epsilon)_1),
\end{eqnarray}
with $H_3=i[H_1,H_2]$ for the unitary $U_{1,3}$.
Following the steps above, we find that
\begin{equation}
 V_J(\theta,H_1,H_2)=U_1(\theta(1+\epsilon_1))U_1(4\epsilon_1\pi\cos(\Phi))\left[1+i\epsilon_1^3M_3(\Phi,H_1,-H_3)+O(\epsilon_1^4)\right]
\end{equation}
Assuming $\epsilon_2$ is small, 
\begin{eqnarray}
\cos(\Phi)&=&\cos(\phi)\cos(\phi\epsilon_2)-\sin(\phi)\sin(\phi\epsilon_2)\nonumber\\
          &\approx&\cos(\phi)-\sin(\phi)\phi\epsilon_2 
\end{eqnarray}
and
\begin{equation}
 V_J(\theta,H_1,H_2)\approx U_1(\theta)\left[1-i\epsilon_1\epsilon_24\pi\phi\sin(\phi)+ i\epsilon_1^3M_3(\Phi,H_1,-H_3)\right].
\end{equation}

The key result is that for fixed $\epsilon_2$ the infidelity scales as $\epsilon_1^6$ when $\epsilon_1$ is large and $\epsilon_2^2\epsilon_1^2$ when $\epsilon_1$ is small. If we can improve the $U_2$ pulses by compensation, we can reduce the error to higher order in $\epsilon_2$. This is exactly how the BB1-WJ sequence (Equation \ref{eqn:VWJ}) is constructed, resulting in a fidelity that scales as $\epsilon_1^6$ when $\epsilon_1$ is large and $\epsilon_1^2\epsilon_2^6$ when $\epsilon_1$ is small.


\section*{References}
\begin{thebibliography}{10}

\bibitem{GottesmanJMO2000}
Gottesman D 2000 {\it J. Mod. Opt} {\bf 47} 333

\bibitem{SvoreQIC07}
Svore K M, DiVincenzo D P and Terhal B M 2007 {\it Quant. Inf. Comp.} {\bf 7} 297

\bibitem{ClarkPRA2009}
Clark C R, Metodi T S, Gasster S D and Brown K R 2009 {\it Phys. Rev.} A {\bf 79} 062314

\bibitem{Freeman:book} 
Freeman R 1999 {\it Spin Choreography} (Oxford: Oxford University Press)

\bibitem{Levitt1986}
Levitt M H 1986 {\it Prog. NMR Spectrosc.} {\bf 18} 61

\bibitem{TyckoPRL1983}
Tycko R 1983 {\it Phys. Rev. Lett.} {\bf 51} 775

\bibitem{BrownPRA2004}
Brown K R, Harrow A W and Chuang I L 2004 {\it Phys. Rev.} A {\bf 70} 052318;
Brown K R, Harrow A W and Chuang I L 2005 {\it Phys. Rev.} A {\bf 72} 039005

\bibitem{XiaoPRA2006}
Xiao L and Jones J A 2006 {\it Phys. Rev.} A {\bf 73} 032334

\bibitem{WimperisJMR1994}
Wimperis S 1994 {\it J. Magn. Reson.} A {\bf 109} 221

\bibitem{JonesPRA2003}
Jones J A 2003 {\it Phys. Rev. A} {\bf 67} 012317

\bibitem{AllwayJMR2007}
Alway W G and Jones J A 2007 {\it J. Magn. Reson.} {\bf 189} 114

\bibitem{HuangJMP1983}
Huang G M, Tarn T J and Clark J W 1983 {\it J. Math. Phys.} {\bf 24} 2608

\bibitem{LiPRA2006}
Li J-S and Khaneja N 2006 {\it Phys. Rev.} A {\bf 73} 030302

\bibitem{ZhangPRA2003}
Zhang J, Vala J, Sastry S and Whaley K B 2003 {\it Phys. Rev.} A {\bf 67} 042313

\bibitem{KhanejaCP2001}
Khaneja N and Glaser S J 2001 {\it Chem. Phys.} {\bf 267} 11

\bibitem{DawsonQIC2006}
Dawson C M and Nielsen M A {\it Quant. Inf. Comp.} {\bf 6} 81

\bibitem{GilchristPRA}
Gilchrist A, Langford N K and Nielsen M A 2005 {\it Phys. Rev.} A {\bf 71} 062310

\bibitem{RakreungdetPRA2009}
Rakreungdet W, Lee J H, Lee K F, Mischuck B E, Montano E and Jessen P S 2009
{\it Phys. Rev.} A {\bf 79} 022316

\bibitem{WeinsteinPRA2005}
Weinstein Y S and Hellberg C S 2005 {\it Phys. Rev.} A {\bf 72} 022319

\bibitem{WeinsteinPRA2005-Levy}
Weinstein Y S, Hellberg C S and Levy J 2005 {\it Phys. Rev.} A {\bf 72} 020304

\bibitem{StorczPRB2005}
Storcz M J, Vala J, Brown K R, Kempe J, Wilhelm F K and Whaley K B 2005 {\it Phys. Rev.} B {\bf 72} 064511

\bibitem{BrownPRA2003}
Brown K R, Vala J and Whaley K B 2003 {\it Phys. Rev.} A {\bf 67} 012309

\bibitem{KempeQIC2001}
Kempe J, Bacon D, DiVincenzo D P and Whaley K B 2001 {\it Quant. Info. Comp.} {\bf 1} 33

\bibitem{KempePRA2002}
Kempe J and Whaley K B 2002 {\it Phys. Rev.} A {\bf 65} 052330

\bibitem{Burkard1999}
Burkard G, Loss D and DiVincenzo D P 1999 {\it Phys. Rev.} B {\bf 59} 2070

\bibitem{Zumbuhl2004}
Zumb\"uhl D M, Marcus C M, Hanson M P and Gossard A C 2004 {\it Phys. Rev. Lett.} {\bf 93} 256801

\bibitem{LidarPRL2001}
Lidar D A and Wu L A 2001 {\it Phys. Rev. Lett.} {\bf 88} 017905

\bibitem{BaconPRL2000}
Bacon D, Kempe J, Lidar D A and Whaley K B 2000 {\it Phys. Rev. Lett.} {\bf 85} 1758

\bibitem{KempePRA2001}
Kempe J, Bacon D, Lidar D A and Whaley K B 2001 {\it Phys. Rev.} A {\bf 63} 042307

\bibitem{DiVincenzoNature2000}
DiVincenzo D P, Bacon D, Kempe J, Burkard G and Whaley K B 2000 {\it Nature} {\bf 408} 339

\bibitem{BaconPRL2001}
Bacon D, Brown K R and Whaley K B 2001 {\it Phys. Rev. Lett.} {\bf 87} 247902

\bibitem{KhanejaJMR2005}
Khaneja N, Reiss T, Kehlet C, Schulte-Herbruggen T and Glaser S J 2005 {\it J. Magn. Reson.} {\bf 172} 296

\bibitem{MagnusCPAM1954}
Magnus W 1954 {\it Commun. Pure Appl. Math.} {\bf 7} 649

\end{thebibliography}
\end{document}